# Driving Skyrmions on a Composite Multiferroic Lattice

Zidong Wang (王子东) and Malcolm J. Grimson
*Department of Physics, the University of Auckland, Auckland 1010, New Zealand*
*E-mail address: Zidong.Wang@auckland.ac.nz*

Magnetic skyrmions and multiferroics are the most interesting objects in nanostructure science that have great potential in future spin-electronic technology. The study of the multiferroic skyrmions has attracted much interest in recent years. This paper reports on the magnetic skyrmions induced by an electric driving field on a composite multiferroic lattice. By using the spin dynamics method, we use a classical magnetic spin model and an electric pseudospin model, which are coupled by a strong magnetoelectric coupling in the dynamical simulations. Interestingly, we observe some skyrmion-like objects on the electric component during the switching process, which is due to the connection between the electric and the magnetic structures.

## I. INTRODUCTION

**A Brief History of Skyrmion:**

The skyrmion is a topological particle-like object, named after Tony H. R. Skyrme, who was firstly described this in quantum field theory in 1960s [1,2]. Decades later with the emergence of spintronics, A. N. Bogdanov *et.al* [3,4,5], successfully predicted that skyrmions can be induced by inhomogeneous chiral interaction, the Dzyaloshinskii-Moriya (DM) interaction, in ferromagnets. The DM interaction [6,7] characterises the asymmetric exchange interaction between the magnetic spin and its neighbours. It is a key ingredient to break the chiral symmetry in magnetic nanostructures [4]. Consequently, the skyrmion has asymmetric spiral-like topological spin texture and it offers numerous advantages for potential spin-electronic technology. One application of magnetic skyrmions, is to provide the low-energy-cost of writing, reading, and erasing non-volatile memory [8].

**Multiferroic Materials:**

Multiferroics offer the possibility of electric-induced-magnetisation and magnetic-induced-(electric) polarisation due to the magnetoelectric coupling between the magnetic dipoles and the electric dipoles [9]. Many experiments [10,11,12,13] have proved the existence of this multiferroism. Furthermore, two types of multiferroic materials have been discovered, the single-phase and the composite-phase [9].

The former, single-phase multiferroics, has received intense investigation in recent years. Seki *et.al*, have observed the magnetic skyrmions controlled by an external electric field on the $Cu_2OSeO_3$ crystal lattice [14]. Generally, it is possible to write the magnetoelectric skyrmions induced by electric polarisation in an insulating chiral multiferroics [15,16]. However, as a single-phase multiferroics, $Cu_2OSeO_3$ has weak magnetic response, and its multiferroism works at a low transition temperature, which is adverse for applications [17].

The other category is the composite multiferroics, which are the artificially synthesized heterostructure of two materials, one a ferromagnet and the other a ferroelectric [18]. Composite multiferroics have remarkable magnetoelectric coupling due to their indirect strain-stress effect [19]. So far, skyrmions have been found in metallic B20 crystals, such as MnSi [20], FeGe [21], $Fe_{1-x}Co_xSi$ [22], and $Mn_{1-x}Fe_xGe$ [23]. Spontaneous skyrmions have been observed in these materials which belong to the chiral ferromagnets [24]. We propose a model of composite multiferroics with a bilayer of a chiral ferromagnetic (FM) layer and a ferroelectric (FE) layer. The magnetoelectric coupling is used to generate multiferroic skyrmions in this work.

**Spin/Pseudospin Models:**

Therefore, a classical spin model can be used to describe the magnetism in the FM structure [25]. In the microscopic origin, the spatial distribution of the magnetisation $M$ is given by a reduced magnetisation $S = M/M_s$, where the saturated magnetisation $M_s$. This is called the magnetic spin.

The dynamics in the composite multiferroics, then involves a microscopic study of electric properties. Generally, the behaviours of the electric polarisation are described by the Landau-Devonshire phenomenological theory [26,27]. But this case, the phenomenological theory has a different length scale to the micromagnetics. Hence we use a pseudospin model to investigate the energy in the FE structure [28,29,30]. This model was introduced by de Gennes [31] and Elliott *et.al* [32]. We have extended it to deal with this system [33,34]. The electric pseudospin $P$ is a polarisation vector. But only the *z*-component of the pseudospins contributes to the energy in the model Hamiltonian.

**Article Outline:**

In this article, we explore the magnetic chiral skyrmions and find electric 'footprint skyrmions' on a composite multiferroic lattice by using stochastic Landau-Lifshitz-Gilbert equations numerically. In Section II, the model of a FM/FE stack structure has been introduced. The spin dynamics method is described in Section III. The electric-field-induced magnetic skyrmions and electric 'footprint skyrmions' are detailed in Section IV. Section V demonstrates the magnetic-field-induced electric 'skyrmions'. The paper concludes with a discussion in Section VI.

## II. MODEL

**Composite Multiferroic Lattice:**

The composite multiferroic lattice has been considered as a two-dimensional FM/FE stack structure, consists of $N \times N$ elements in each layer. The schematic view is in Fig. 1, with the magnetic spins (red arrows) and the electric pseudospins (blue arrows). The FM structure and the FE structure are glued together by the magnetoelectric coupling. Note that, in this numerical simulation each magnetic spin coupled with an electric pseudospin.

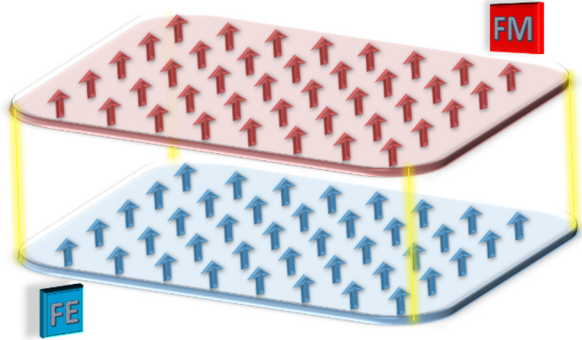

Fig. 1. Schematic illustration of a FM/FE composite multiferroic lattice. The top-plane represents the FM structure with magnetic spins as red arrows; the bottom-plane represents the FE structure with electric pseudospins as blue arrows.

Therefore, the total energy $\mathcal{H}$ for the microscopic model can be written as a sum of three terms; the Hamiltonian in the FM structure $\mathcal{H}_{FM}$, the Hamiltonian in the FE structure $\mathcal{H}_{FE}$ and the magnetoelectric interaction between the FM and the FE structure $\mathcal{H}_{ME}$.

$$\mathcal{H} = \mathcal{H}_{FM} + \mathcal{H}_{FE} + \mathcal{H}_{ME} \quad (1)$$

**Energy in the FM Structure (classical Heisenberg model):**

The spin system of FM structure can be described by a classical Heisenberg model on the two-dimensional square lattice. The magnetic spin represents as $\mathbf{S}_{i,j} = \left(S_{i,j}^x, S_{i,j}^y, S_{i,j}^z\right)$, is a normalized operator with unit size, i.e. $\|\mathbf{S}_{i,j}\| = 1$, and $i,j \in [1,2,3,\ldots N]$ defined the location of the spin. The Hamiltonian $\mathcal{H}_{FM}$ is given by:

$$\mathcal{H}_{FM} = \mathcal{H}_{FM}^{int} + \mathcal{H}_{FM}^{dmi} + \mathcal{H}_{FM}^{ani} + \mathcal{H}_{FM}^{ext} \quad (2)$$

The first term stands for the nearest-neighbour exchange interaction, and $J_{FM}^* = J_{FM}/k_B T$ is the dimensionless exchange interaction coupling coefficient.

$$\mathcal{H}_{FM}^{int} = -J_{FM} \sum_{i,j} \left[\mathbf{S}_{i,j} \cdot \left(\mathbf{S}_{i+1,j} + \mathbf{S}_{i,j+1}\right)\right] \quad (3)$$

The second term stands for the DM interaction, which is a non-linear exchange interaction that specifies the helicity of the skyrmions.

$$\mathcal{H}_{FM}^{dmi} = -D_{FM} \sum_{i,j} \left[\mathbf{S}_{i,j} \times \mathbf{S}_{i+1,j} \cdot \hat{x} + \mathbf{S}_{i,j} \times \mathbf{S}_{i,j+1} \cdot \hat{y}\right] \quad (4)$$

In Eq. (4), $D_{FM}^* = D_{FM}/k_B T$ is the dimensionless DM interaction coefficient, $\hat{x}$ and $\hat{y}$ are the unit vectors of the $x$- and $y$-axes respectively. The third term stands for the magnetic anisotropy.

$$\mathcal{H}_{FM}^{ani} = -K_{FM}^z \sum_{i,j} \left(S_{i,j}^z\right)^2 \quad (5)$$

In Eq. (5), $K_{FM}^* = K_{FM}^z/k_B T$ is the dimensionless uniaxial anisotropic coefficient in the $z$-direction. The fourth term stands for the external Zeeman energy.

$$\mathcal{H}_{FM}^{ext} = -\mu_0 \chi_m H_{ext}^z(t) \sum_{i,j} S_{i,j}^z \quad (6)$$

In Eq. (6), $H_{ext}^*(t) = \mu_0 \chi_m H_{ext}^z(t)/k_B T$ is a dimensionless external time-depended magnetic field, is applied perpendicular to the lattice sample along the $z$-direction, $\mu_0$ is the magnetic permeability of the classical vacuum, and $\chi_m$ is the magnetic susceptibility in the FM materials.

**Energy in the FE Structure (transverse Ising model):**

We have studied the electric dipoles in the system by the transverse Ising model (TIM) for the electric pseudospins [33,34]. The electric pseudospin on the FE structure are regard as a vector $\mathbf{P}_{k,l} = (P_{k,l}^x, P_{k,l}^y, P_{k,l}^z)$, and $k,l \in [1,2,3,\ldots N]$ characterises their locations. The Hamiltonian $\mathcal{H}_{FE}$ of the electric subsystem is modelled by a transverse Ising model, is given by:

$$\mathcal{H}_{FE} = \mathcal{H}_{FE}^{int} + \mathcal{H}_{FE}^{tran} + \mathcal{H}_{FE}^{ext} \quad (7)$$

In the transverse Ising model, only the $z$-component of the electric pseudospins has a contribution to the Hamiltonian, and $J_{FE}^* = J_{FE}/k_B T$ is the dimensionless nearest-neighbour interaction coefficient between the electric pseudospins.

$$\mathcal{H}_{FE}^{int} = -J_{FE} \sum_{k,l} \left[P_{k,l}^z \left(P_{k+1,l}^z + P_{k,l+1}^z\right)\right] \quad (8)$$

The second term stands for the transverse energy [35], where $\Omega_{FE}^* = \Omega_{FE}^x/k_B T$ is a dimensionless transverse field in the $x$-direction, which is perpendicular to the Ising $z$-direction [36].

$$\mathcal{H}_{FE}^{tran} = -\Omega_{FE}^x \sum_{k,l} P_{k,l}^x \quad (9)$$

The third term stands for the external energy provided by an additional electric driving field.

$$\mathcal{H}_{FE}^{ext} = -\epsilon_0 \chi_e E_{ext}^z(t) \sum_{k,l} P_{k,l}^z \quad (10)$$

In Eq. (10), $E_{ext}^*(t) = \epsilon_0 \chi_e E_{ext}^z(t)/k_B T$ is an dimensionless external time-dependent electric field, applied perpendicular to the lattice sample along the $z$-direction, $\epsilon_0$ is the electric permittivity of free space, and $\chi_e$ is the dielectric susceptibility.

The size of an electric pseudospin is different from the classical magnetic spin. The polarisation is defined as the electric dipole moment density in dielectric materials. The dipole moment density P is proportional to the external electric field $E_{ext}$ [37], with

$$P = \epsilon_0 \chi_e E_{ext} \quad (11)$$

In the spin system, the size of each electric pseudospin is proportional to the magnitude of its effective field $\left\|E_{P_{k,l}}^{eff}\right\|$ [4], with

$$\left\|\boldsymbol{P}_{k,l}\right\| = \epsilon_0 \aleph_e \left\|\boldsymbol{E}_{P_{k,l}}^{eff}\right\| \quad (12)$$

where $\aleph_e$ is the dimensionless pseudoscalar susceptibility. In consequence, the electric pseudospins have variable size like the behaviour of the electric dipoles.

**Magnetoelectric Interactions:**
The behaviour of the multiferroic is related to the magnetoelectric coupling at the interface and this can be described by the spin-dipole interaction Hamiltonian $\mathcal{H}_{ME}$ [9].

$$\mathcal{H}_{ME} = -g_m \sum_m \sum_{i,j,k,l} \left(S_{i,j}^z P_{k,l}^z\right)^m \quad (13)$$

In Eq. (13), $g_m^* = g_m/k_B T$ is the dimensionless magnetoelectric coupling coefficient, and $m$ is the energy excitation index [38].

Note that, we only need to account for the low-energy excitations between the layers and so we restrict ourselves to the linear term (i.e. $m=1$). Higher order terms have not been studied here, for simplicity and due to their minor effects in the numerical modelling.

## III. METHOD

The time evolution of the spins/pseudospins responses is studied by numerically solving the Landau-Lifshitz-Gilbert equations. The role of the thermal effects have been considered by including as a Gaussian white noise contribution.

**Dynamics of Magnetic Spins:**

In the FM structure, Eq. (14) shows the differential equation which predicts the rotation of a magnetic spin in response to its torques,

$$\frac{\partial \boldsymbol{S}_{i,j}}{\partial t} = -\gamma_{FM}\left[\boldsymbol{S}_{i,j} \times \boldsymbol{H}_{S_{i,j}}^{eff}\right] - \lambda_{FM}\left[\boldsymbol{S}_{i,j} \times \left(\boldsymbol{S}_{i,j} \times \boldsymbol{H}_{S_{i,j}}^{eff}\right)\right]$$
(14)

where $\gamma_{FM}$ is the gyromagnetic ratio which relates the magnetic spin to its angular momentum, $\lambda_{FM}$ is the phenomenological Gilbert damping term in the FM structure, and $\boldsymbol{H}_{S_{i,j}}^{eff}$ is the effective field of each magnetic spin. This is the derivative of the system Hamiltonian of Eq. (2) with respect to the magnitudes of the magnetic spin in each direction, with:

$$\boldsymbol{H}_{S_{i,j}}^{eff} = -\frac{\delta \mathcal{H}}{\delta \boldsymbol{S}_{i,j}} + \boldsymbol{h}_{S_{i,j}}^r \quad (15)$$

where $\boldsymbol{h}_{S_{i,j}}^r = \left(h_{S_{i,j}}^x, h_{S_{i,j}}^y, h_{S_{i,j}}^z\right)$ is the three-dimensional stochastic field acting on each magnetic spin.

**Dynamics of Electric Pseudospins:**
In the FE structure, the pseudospins describe the locations of the electric dipole. The electric dipole moment is a measure of the separation of positive and negative charges in $z$-direction, and it is a scalar. Consequently, the time evolution of the electric pseudospin is expected to perform a precession free trajectory [28,33,34], with

$$\frac{\partial \boldsymbol{P}_{i,j}}{\partial t} = -\lambda_{FE}\left[\boldsymbol{P}_{i,j} \times \left(\boldsymbol{P}_{i,j} \times \boldsymbol{E}_{P_{i,j}}^{eff}\right)\right] \quad (16)$$

where $\lambda_{FE}$ is the phenomenological Gilbert damping term in the FE structure, and $\boldsymbol{E}_{P_{i,j}}^{eff}$ is the electric effective field for each pseudospin. It is defined as a functional derivative of Eq. (3), is given as:

$$\boldsymbol{E}_{P_{k,l}}^{eff} = -\frac{\delta \mathcal{H}}{\delta \boldsymbol{P}_{k,l}} + \boldsymbol{h}_{P_{k,l}}^r \quad (17)$$

where $\boldsymbol{h}_{P_{l,k}}^r = \left(h_{P_{k,l}}^x, h_{P_{k,l}}^y, h_{P_{k,l}}^z\right)$ is the three-dimensional stochastic field act on each electric pseudospin. Note that, only the $z$-component of the pseudospin represents the electric polarisation.

**Thermal Effect:**
The thermal influence cannot be neglected, and here, we introduce a stochastic field $\boldsymbol{h}^r$, which is a Gaussian white noise, into the dynamics [39]. In Eq. (18), $\mu$ is the mean of Gaussian distribution, $T^*$ is the dimensionless Boltzmann temperature represents the standard deviation of distribution, which is proportional to the temperature. $\Delta \boldsymbol{r}$ is the random variable vector, which has three degrees of freedom, i.e. $\Delta \boldsymbol{r} = (\Delta x, \Delta y, \Delta z)$.

$$h^r = \frac{1}{T^*\sqrt{2\pi}} e^{\frac{-(\Delta r - \mu)^2}{2(T^*)^2}} \quad (18)$$

Thus, Eqs. (14) and (16) become stochastic Landau-Lifshitz-Gilbert equations. In this work, we shall assume a cool thermal background, in order to clearly show the behaviour of FM and FE. Hence, the Gaussian distribution should have no bias, i.e. $\mu = 0$, and the Boltzmann temperature $T^* = 0.01$ are used in all numerical simulations below.

## IV. SKYRMIONS BY AN ELECTRIC FIELD

The stochastic Landau-Lifshitz-Gilbert equations are solved by a fourth-order Range-Kutta method with a time step for $\Delta t = 0.0001$. We implement a dimensionless parameter set: { $J_{FM}^* = 1$, $D_{FM}^* = 1$, $K_{FM}^* = 0.1$, $J_{FE}^* = 1$, $\Omega_{FE}^* = 0.1$, $\aleph_e^* = 0.1$, $g_1^* = 0.5$, $\gamma_{FM}^* = 1$ and $\lambda_{FM}^* = \lambda_{FE}^* = 0.1$ } in this section. Note that '*'s characterise dimensionless quantities in this article. The number of magnetic spins and electric pseudospins $N_{FM} = N_{FE} = 23 \times 23$ are used. Free boundary conditions and random initial state are applied. The FM/FE stack lattice is driven by the electric field only (i.e. no magnetic external field $H_{ext}^*(t) = 0$).

**Magnetic Chiral Skyrmions:**

We obtain the chiral skyrmions induced by a square electric field with a dimensionless amplitude $E_0^* = 10$ on the FM structure. The result of the first half square wave (i.e., applied electric field $E_{ext}^*(t) = 10$ and $t \in [0,310]$) is shown in Fig. 2. The nets magnetisation (red curve) and electric polarisation (blue curve) are depicted in Fig. 2(a). Four snapshots show the dynamical progress of the magnetic skyrmion creation in Figs. 2(b) → 2(c) → 2(d) → 2(e). Initially, electric pseudospins have been quickly polarised by the electric driven field [Fig. 2(b) with Time $("\Delta")=1$]. A short time later, the FE structure has been completely well-aligned, and the FM structure starts to order [Fig. 2(c) with Time $("*")=5$]. 'Baby' skyrmions can be observed in Fig. 2(d) with Time $("X")=20$. The 'baby' skyrmions are formed around the edge have very short life, and eventually, five of them in the bulk lattice survive as stabilised chiral skyrmions [Fig. 2(e) with Time $("O")=300$]. The stabilised skyrmions require the entire system has lowest free energy. This can be manipulated by the alignment and the size of these skyrmions.

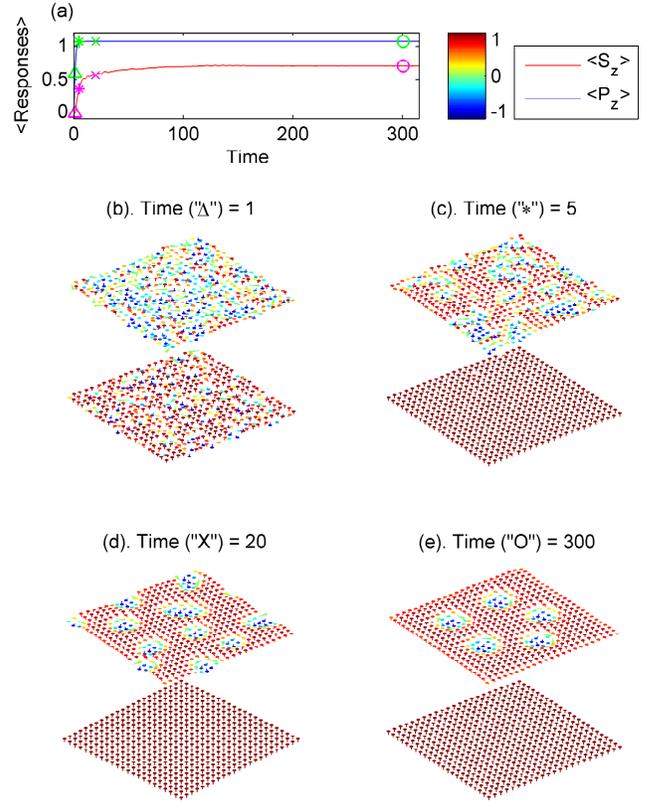

Fig. 2. (a) Net responses of the magnetic spins (red) and the electric pseudospins (blue) to an electric field for $E_{ext}^*(t) = 10$ and $t \in [0,310]$. Four moments are selected with symbols "$\Delta$", "$*$", "X" and "O" labelled in the panel. (b) – (e) show the real-time details of these four moments, respectively. The top-plane is the FM lattice, bottom-plane is the FE lattice. The colour scale represents the magnitude of the *z*-component.

Importantly, the magnetoelectric coupling in the composite multiferroics acts as the source of a magnetic driving field in the chiral magnets. Since each magnetic spin is bound with an electric pseudospin, the magnetic skyrmions can be manipulated by their relativistic electric polarisation. As a consequence, the second half of the square electric wave points to opposite direction (i.e. $E_{ext}^*(t) = -10$), and the reversed skyrmions are formed on the FM structure, as shown in Fig. 3.

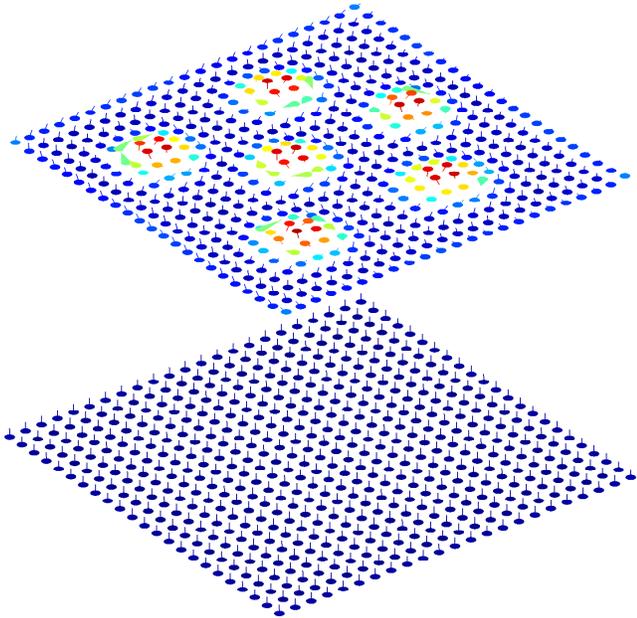

Fig. 3. Stabilised magnetic skyrmions driven by a negative electric driving field $E_{ext}^*(t) = -10$. The colour scales for the FM structure (top-plane) and the FE structure (bottom-plane) are corresponding to the colour bar in Fig. 1.

**Electric 'Footprint Skyrmions':**

Interestingly, during the switching process (which introduces as a sudden change of the direction of the electric field in the present example, i.e. $E_{ext}^*(t) = 10 \to -10$ ), we observe some skyrmion-like features appearing on the FE structure, as shown in Fig 4. Fig. 4(a) presents the net magnetisation (red curve) and the net electric polarisation (blue curve) in a limited time of $t \in [316, 326]$, which is taken around the switching of the square electric driving field. As the symbols ("$\Delta$", "$*$", "X" and "O") seen in Fig. 4(a), we provide snapshots of these four times in Figs. 4(b) – (e). The top-plane is the FM lattice with magnetic chiral skyrmions, and the bottom-plane is the FE lattice. The magnetic skyrmions are consistent with the skyrmion footprint on the FE lattice. Hence, we call them here as the *electric 'footprint skyrmions'*. They only appear for an extremely short period during the reorienting process, as shown in Figs. 4(b) $\to$ 4(c) $\to$ 4(d) $\to$ 4(e).

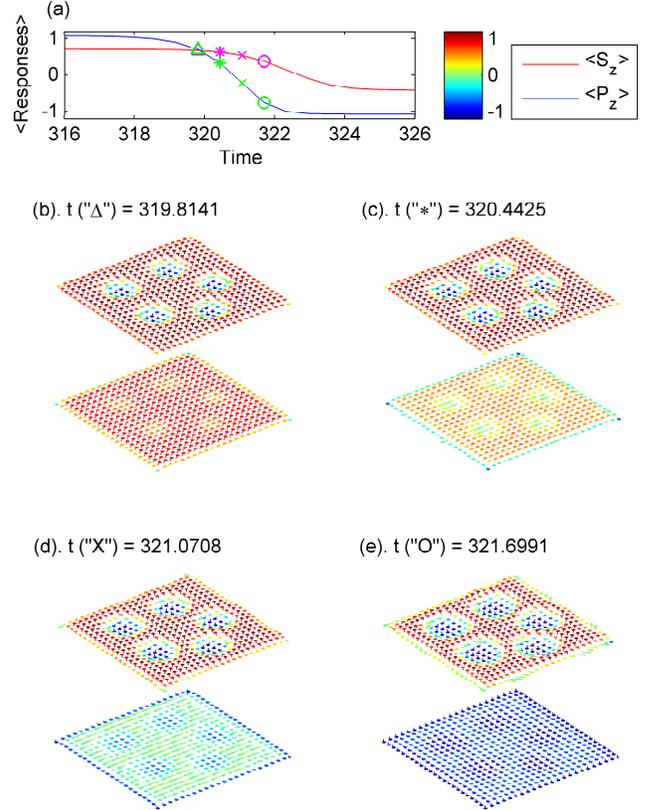

Fig. 4. Electric 'footprint skyrmions' during the switching process (i.e. $E_{ext}^*(t) = 10 \to -10$ dimensionless). Same details as Fig. 2.

To understand why the electric 'footprint skyrmions' can emerge, we consider the *footprint effect*. During the switching process, the exchange energy is used to support electric pseudospins responding to the driving field. Therefore, as the reducing of the exchange energy, the binding energy (magnetoelectric interaction) dominates the system. The magnetic skyrmions provide non-uniform magnetisation distribution in the FM structure. This indicates that different binding energies in each spin-pseudospin bond result the 'footprints' on the FE structure. These 'footprints' exist for a limited time, since the electric pseudospins reorient quickly to respond to the driving field. Eventually, since the exchange energy has been saturated, the 'footprints' disappear, as shown in Fig. 4(e). Consequently, we can control the electric 'footprint skyrmions' by the driving-field application. To confirm this, in next section we demonstrate that the stabilised electric 'footprint skyrmions' can be induced by an external magnetic field.

# V. ELECTRIC 'FOOTPRINT SKYRMIONS' BY A MAGNETIC FIELD

In this work, one of our aims is to generate the skyrmion-like features in the electric materials. In Section IV, the so-called electric 'footprint skyrmions' induced by the electric field have been numerically observed during the switching process. Now, we replace the square electric driving field by a different static magnetic driving field, in order to enhance the stability of electric 'footprint skyrmions'. A dimensionless parameter set: { $J^*_{FM}=1$, $D^*_{FM}=1$, $K^*_{FM}=0.1$, $J^*_{FE}=0.01$, $\Omega^*_{FE}=0.1$, $\aleph^*_e=1$, $g^*_1=0.5$, $\gamma^*_{FM}=1$ and $\lambda^*_{FM}=\lambda^*_{FE}=0.1$ } is selected to study the magnetic-field-driven dynamics with the same lattice sample $N_{FM}=N_{FE}=23\times 23$, free boundary conditions, and a static magnetic field $H^*_{ext}(t)=0.5$. Note that, we use a dielectric (weak) exchange coupling coefficient $J^*_{FE}=0.01$ in this section.

**Magnetic-Field-Induced Electric 'Footprint Skyrmion':**

Fig. 5 demonstrates an electric 'footprint skyrmion' can also be induced by a magnetic driving field on the stack FM/electric lattice. A magnetic skyrmion is formed on the chiral FM structure (top-plane) due to its direct response to the magnetic driving field. Since the electric pseudospin (bottom-plane) is coupled with its relativistic magnetic spin by the magnetoelectric effect, hence a 'footprint skyrmion' is projected on the electric structure. Importantly, the electric 'footprint skyrmion' does not have the spiral texture. This is seen in the bottom-plane in Fig. 5. The pseudospins represent electric dipoles, and the electric dipoles only have magnitudes, but no directions. Therefore, the electric pseudospins located in the 'footprint skyrmion' have different magnitudes to pseudospins elsewhere. In Fig. 5 bottom-plane, this can be observed by the different colours. In the absence of electric external field and with a weak exchange interaction, the sizes of pseudospins are finite [see Eq. (12)]. Thus the colour in the electric structure is lighter than the FM structure. This type of electric 'footprint skyrmions' is consistent with the corresponding magnetic skyrmions.

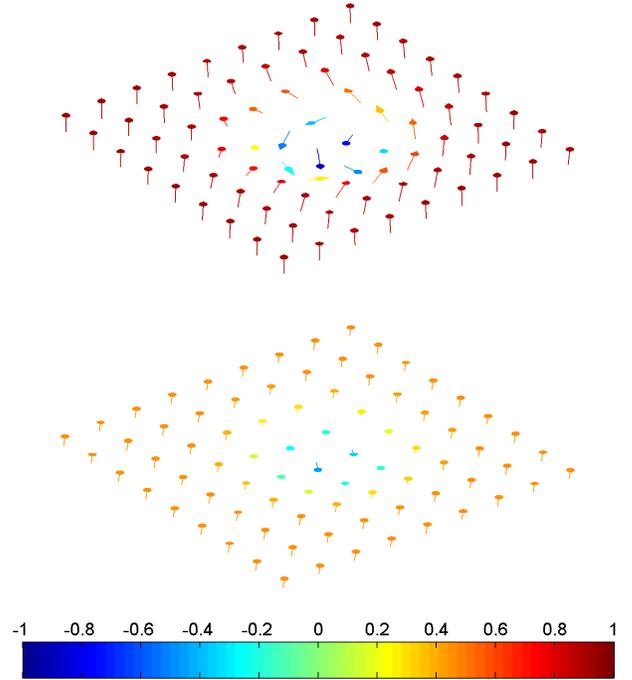

Fig. 5. A magnetic skyrmion (top-plane) on the FM structure, and its relativistic electric 'footprint skyrmion' (bottom-plane) on the electric structure, to an applied magnetic field $H^*_{ext}(t)=0.5$ (dimensionless). The colour scales represent the magnitudes of the *z*-component.

**Effects of the Exchange Coupling in the electric Structure:**

In Section IV, we have discussed that the electric 'footprint skyrmions' can be created during the switching process, due to the magnetoelectric interaction dominates the energy in the FE structure. It is the key feature to obtain the electric 'footprint skyrmions'. In this section, the electric external field is absent, and the transverse energy is small. Thus, the exchange interaction only competes with the magnetoelectric interaction. The exchange interaction is modulated by the exchange coupling $J_{FE}$ in the simulation. In Fig. 6, three results are shown for different electric exchange coupling coefficients ($J^*_{FE}=0.01$, $J^*_{FE}=0.1$, and $J^*_{FE}=0.2$), and we compare the skyrmions under a static magnetic field $H^*_{ext}(t)=0.5$. In Fig. 6(a) with $J^*_{FE}=0.01$, it contains four magnetic skyrmions with the electric 'footprint skyrmions'. Later on, as $J_{FE}$ increases, the skyrmions reduce their size and number. As seen in Fig 6(b) with $J^*_{FE}=0.1$, two tiny skyrmions and two 'footprint skyrmions' have survived. Finally, neither skyrmions nor 'footprint skyrmions' survive as observed in Fig. 6(c) with $J^*_{FE}=0.2$. The latter can be traced back to the coupling between the FM structure and electric structure by the magnetoelectric effects.

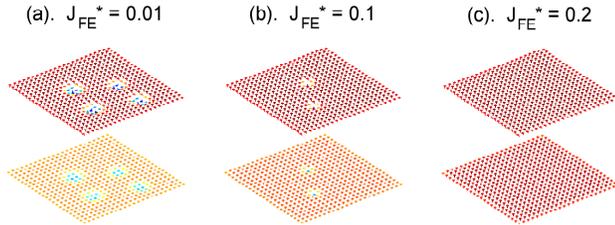

Fig. 6. A comparison of the electric exchange coupling effects for the skyrmions on the composite multiferroic lattice. The colour scales for the magnetic spins (top-plane) and the electric pseudospins (bottom-plane) are corresponding to the colour bar in Fig. 5.

## VI. CONCLUSION

We have demonstrated that the magnetic skyrmions on the chiral FM structure can be driven by an electric field. Interestingly, since the FM and FE structures are coupled by the magnetoelectric effects, it offers an opportunity for magnetic skyrmions on the FM structure to produce projections onto the FE structure. We call these projections 'footprint skyrmions'. The electric 'footprint skyrmions' can be generated by either an electric field, or a magnetic field. In the electric driving field, the 'footprint skyrmions' only exist during the switching of the electric field. In the magnetic driving field, the 'footprint skyrmions' are stable.

## ACKNOWLEDGEMENTS

Z. D. Wang gratefully acknowledges Zhao BingJin (赵秉金), Wang YuHua (王玉华), Zhao WenXia (赵雯霞) and Wang Feng (王峰) for financial support.